\documentclass[journal]{IEEEtran}
\IEEEoverridecommandlockouts
\usepackage{amsmath,amsfonts}
\usepackage{amsmath}
\usepackage{algorithm}
\usepackage{algpseudocode}
\usepackage{array}
\usepackage[caption=false,font=normalsize,labelfont=sf,textfont=sf]{subfig}
\usepackage{textcomp}
\usepackage{stfloats}
\usepackage{url}
\usepackage{verbatim}
\usepackage{graphicx}
\usepackage{cite}
\usepackage{xcolor}
\usepackage{subfig}
\usepackage{booktabs}
\begin{document}

\title{Robust Channel Estimation for Optical Wireless Communications Using Neural Network}

\author{\IEEEauthorblockN{Dianxin Luan,~\IEEEmembership{Graduate Student Member,~IEEE,} John Thompson,~\IEEEmembership{Fellow,~IEEE}}\\ 
\IEEEauthorblockA{\textit{Institute for Imaging, Data and Communications, School of Engineering, University of Edinburgh}}
}

\maketitle

\begin{abstract}
Optical Wireless Communication (OWC) has gained significant attention due to its high-speed data transmission and throughput. Optical wireless channels are often assumed to be flat, but this paper considers very dispersive optical wireless environments resulting in frequency-selective scenarios. To address this, this paper presents a robust and low-complexity channel estimation framework to mitigate frequency-selective effects, then to improve system reliability. This channel estimation framework contains a neural network with strong generalization to provide estimated information about the environment. Based on this estimate and the corresponding delay spread, the selector will activate one of several candidate neural networks to precisely predict this channel. Simulation results demonstrate that the proposed method has improved and robust normalized mean square error (NMSE) and bit error rate (BER) performance in dynamic environments. These findings highlight the potential of extending high-data rate and reliable communications under indoor multi-tap optical channels. 
\end{abstract}


\section{Introduction}
\IEEEPARstart{W}{ireless} communications are critical for current applications but largely rely on radio frequency (RF) systems. Compared to RF, optical wireless communications (OWC) can offer higher data rates, inherent security and lower transmission power. This ensures OWC is a competitive solution that emerge as a leading technology for next-generation wireless networks to offer high bandwidth and inherent security \cite{wang2023road}, including visible-light communication (VLC), light fidelity (LiFi) and free-space optical (FSO) links, in diverse applications, from indoor networks to underwater communications \cite{amran2023link}. However, the performance of wireless systems depends on an accurate estimate of the state of the channel. Conventional methods such as least-squares (LS) and minimum mean square error (MMSE) are widely applied for RF applications, and neural networks have been investigated to improve performance and reliability \cite{luan2022attention, luan2023channelformer, liu2024pd}. 

The line-of-sight (LOS) component is often tens of decibels stronger than the non-line-of-sight (NLOS) components arising from reflections off walls, ceilings and other surfaces as described in IEEE 802.11bb document, leading to a common flat-fading assumption. The paper \cite{chen2025adaptive} proposes a camera-based channel estimation method to estimate attenuation, adapt modulation and power accordingly. For neural network based methods, a joint frame detection and channel estimation is proposed in \cite{jiang2017joint} for Direct Current-biased Optical OFDM (DCO-OFDM) LiFi systems, achieving good synchronization under LED-induced ISI and superior performance. The work \cite{elfikky2023symbol} presents a neural network-based channel estimation framework for space optical communication systems, which demonstrates superior performance and achieves good accuracy under diverse turbulence-induced fading scenarios such as Log-normal and Gamma-Gamma channels. However, despite the characteristics of optical wireless channels, multi-path propagation can still give rise to frequency-selective fading which requires robust channel estimation techniques. 

In this paper, we propose an adaptive neural network solution to estimate wireless optical channels robustly and reduce the bit errors at the receiver end. According to pre-estimates of optical channels, this adaptive solution activates a suitable neural network to achieve overall prediction. For simulations, we employ a ray tracing simulator for realistic modeling of the environment, which shows frequency-selective scenarios for orthogonal frequency-division multiplexing (OFDM) waveforms, even when the reflection path has only a very small gain. The code for this deployed OWC channel code is available at https://github.com/dianixn/OWC-Channels. 
\section{System settings and channel model}
\label{System settings}
This paper considers an uncoded DCO-OFDM system for indoor communications, where intensity modulation and direct detection (IM-DD) of the optical carrier using an incoherent light source is implemented. Each slot consists of $N_f=324$ subcarriers and $N_s$=14 OFDM symbols. The frequency-domain allocation for the demodulation reference signal comprises 4 pilot symbols which are the 3\textsuperscript{rd}, 6\textsuperscript{th}, 9\textsuperscript{th} and 12\textsuperscript{th} OFDM symbols. For each pilot OFDM symbol, the indices of the pilot subcarriers start from the first subcarrier and are spaced by $L_s=5$ subcarriers and the remaining subcarriers are set to 0 for the first half subcarriers. All of the data subcarriers in the data OFDM symbols are assigned 64 Quadrature Amplitude Modulation (64-QAM) modulated symbols. For each OFDM symbol $s(i)$, complex baseband signals are constrained to have Hermitian symmetry, i.e. $\forall i \in \left[0, N_f - 1\right], s(i) = s^{*}(N_f - i)$ for the second half subcarriers. The frequency-domain OFDM symbols are then processed by the inverse fast Fourier transform (IFFT). The cyclic prefix (CP) with length of $L_{CP}=7$ is inserted and a DC bias is applied to ensure the transmitted signal is always positive. 
This paper considers OWC channel that retains constant for each slot, which is given by 
\begin{equation}
    g(t) = h_{\mathrm{LOS}}\delta\left(t\right) + \sum_{m=1}^{M-1} h_{\mathrm{NLOS}}^{m}\delta\left(t-\tau_mT_s\right), 
\label{channel impulse response}
\end{equation}
where $h_{\mathrm{LOS}}$ is the LOS path gain and $h_{\mathrm{NLOS}}^{m}$ is the NLOS path gain for the $m$th path which are real and nonnegative. $\tau_m$ is the reflection delay normalized by the sampling period $T_s$ and $\delta\left(\cdot\right)$ is the Dirac delta function. This paper only takes into account specular reflections because diffuse reflections have a much smaller path power. The LOS and NLOS links are given by \cite{barry1993simulation, mmbaga2016performance} 
\begin{equation}
    h_{\mathrm{LOS}} = \begin{cases}
    \frac{A((k+1)\cos^{k}(\phi)\cos(\theta)}{2\pi d^2}, & \theta \in [0, \varphi_{\frac{1}{2}}] \\ 
    0, & \text{otherwise} \\
\end{cases} , 
\end{equation}
\begin{equation}
    h_{\mathrm{NLOS}} = \alpha \frac{A((k+1)\cos^{k}(\phi_{Tx})\cos(\theta_{Tx})}{2\pi d_{Tx}^{2}}, 
\end{equation}
where $k = \frac{-\mathrm{ln}(2)}{\mathrm{ln}\mathrm{cos}(\Phi_{\frac{1}{2}})}$. $A$ is the collection area of the photodiode (PD), $\Phi_{\frac{1}{2}}$ is the transmitter semi-angle, $\varphi_{\frac{1}{2}}$ is the FOV semi-angle of the receiver and $\alpha$ is the surface reflection coefficient of the wall. $\theta, \phi$ is the angle of incidence and emergence with respect to the receiver detector direction, and $d$ is the distance between the transmitter and the receiver. For the specular reflection path, $\theta_{Tx}, \phi_{Tx}$ is the angle of incidence and emergence from the image position of the transmitter to the receiver detector direction, $d_{Tx}$ is the distance between the transmitter image and the receiver. By removing the CP and applying FFT operation, the received signal $\mathbf{Y} \in \mathbb{C}^{{N_f}\times N_s}$ is 
\begin{equation}
\mathbf{Y} = \mathbf{H} \circ \mathbf{X} + \mathbf{W}, 
\end{equation}
where $\mathbf{X}, \mathbf{H} \in \mathbb{C}^{{N_f}\times N_s}$ are the frequency domain transmitted signal and the channel matrix. $\mathbf{W} \in \mathbb{C}^{{N_f}\times N_s}$ denotes the sum of ambient light shot and thermal noises with a variance of $\delta^{2} = \delta_{\mathrm{slot}}^{2} + \delta_{\mathrm{thermal}}^{2}$. The operator $\circ$ represents the Hadamard product. The received pilot signal is extracted to provide a channel reference for the complete packet. The LS estimate is calculated by $\mathbf{\hat{H}^{LS}} = \mathbf{Y_p}\oslash\mathbf{X_p}$ where $\mathbf{Y_p}, \mathbf{X_p} \in \mathbb{C}^{\frac{N_f}{L_s}\times 4}$ are the received and transmitted pilot signals respectively, and $\oslash$ is the Hadamard division. To reduce the mean square error (MSE) of this estimate, the linear MMSE solution is \cite{edfors1998ofdm} 
\begin{equation}
\mathbf{\hat{H}^{MMSE}} = \mathbf{R_{HH_{p}}}\left(\mathbf{R_{H_{p}H_{p}}} + \delta^2 \mathbf{I}\right)^{-1}\mathbf{\hat{H}^{LS}}, 
\end{equation}
where $\mathbf{R_{HH_{p}}}$ is the cross-correlation matrix between the actual channel matrix and the actual channel matrix at the pilot positions, $\mathbf{R_{H_{p}H_{p}}}$ is the autocorrelation matrix of $\mathbf{H_p}$ and $\mathbf{I}$ is the identity matrix. After equalization, the processed signal is decoded to predict the bit-level data. 
\section{Robust channel estimation for wireless optical channels}
\label{Robust channel estimation}
\begin{figure}[htbp]
\centerline{\includegraphics[width=0.39\textwidth]{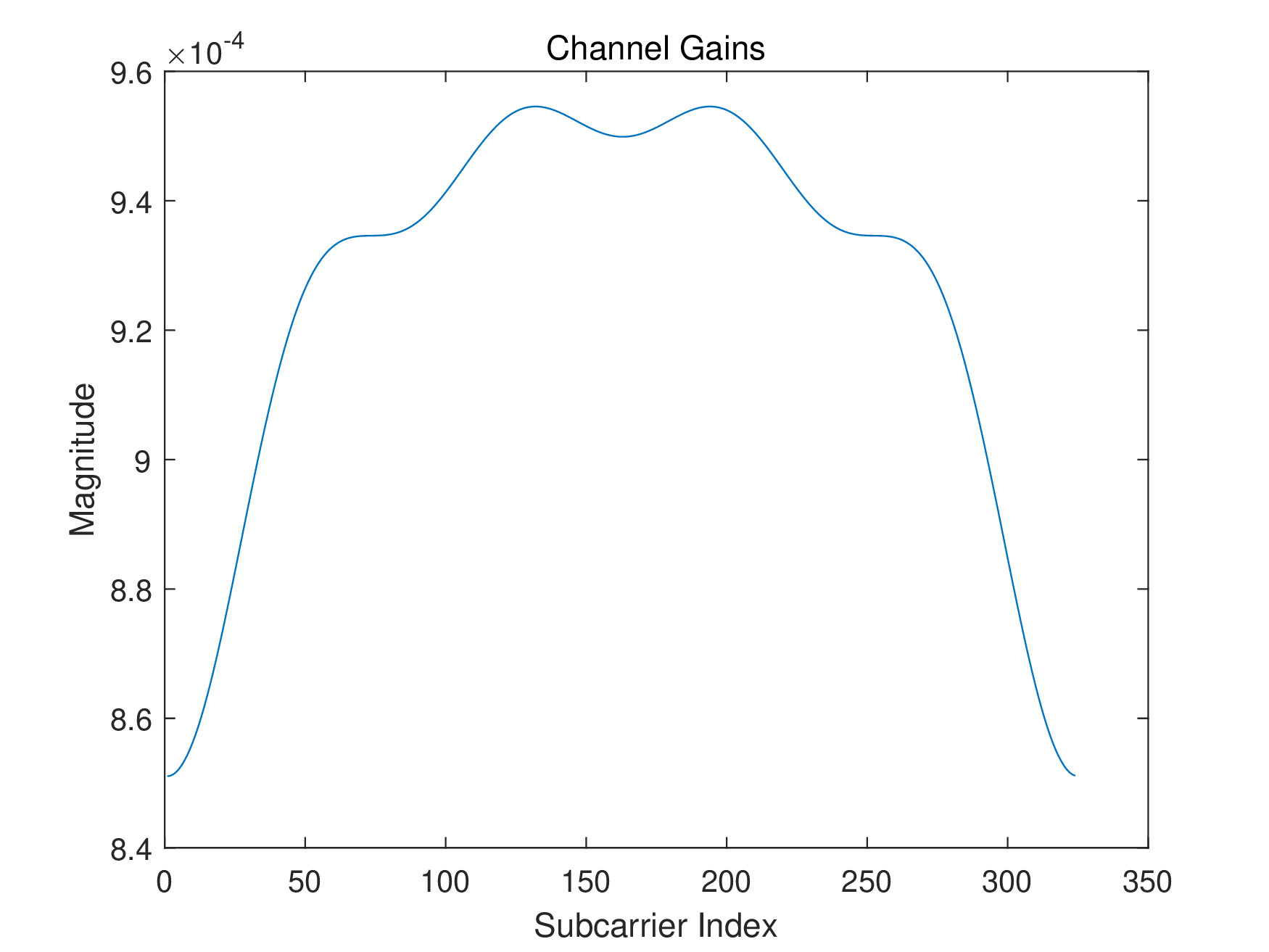}}
\caption{Example of a two-path optical wireless channel where the magnitude of the reflection path is tens dB less than that of the LOS path, but channel gain varies among these subcarriers. }
\label{multitap}
\end{figure}

Typically the optical wireless channel is assumed to be flat, but the sampled channel impulse response still possibly has multiple taps for a OFDM waveform. Transferring equ.~(\ref{channel impulse response}) into the frequency domain and then sampling this continuous expression, the $k$th channel gain and the corresponding channel impulse response are shown above. Equ.~(\ref{channel tap}) shows that the sampled channel impulse response exhibits time dispersion characteristics due to the reflection path. This possibly results in weak frequency selective scenarios as shown in Fig.~\ref{multitap}. 
\begin{figure*}
\begin{equation}
\mathbf{H}_k = h_{\mathrm{LOS}} + \sum_{m=1}^{M-1} h_{\mathrm{NLOS}}\frac{e^{-j\frac{2\pi}{N_f}k\left(\tau_m-\frac{N_f-1}{2}\right)}\sin\left(\frac{\pi}{N_f}\bigl(2N_f\tau_m-k\bigr)\right)}{\sin\left(\frac{\pi}{N_f}k\right)}, 
\label{H_K}
\end{equation}

\begin{equation}
    \mathbf{h}(n) = h_{\mathrm{LOS}}\delta\left(n\right) + \sum_{m=1}^{M-1} h_{\mathrm{NLOS}}\frac{\left(\sin\left( \frac{\pi}{N_f} \left(n + (2N_f - 1)\tau_m \right) \right) - \sin\left( \frac{\pi}{N_f} (n - \tau_m) \right) \right)}{2\sin(\frac{\pi}{N_f}(\tau_m-n))}. 
\label{channel tap}
\end{equation}
\end{figure*}

To address this, a novel channel estimator is required to compensate for this effect. In this paper, we propose a multi-branch but non-iterative neural estimation framework that adaptively selects the most appropriate neural estimator based on the estimated delay spread of the optical channel. Each branch contains a neural network that adapt to optical channels with low delay spreads (LDS), medium delay spreads (MDS) or high delay spreads (HDS). This paper employs InterpolateNet \cite{luan2021low} as an example neural network. It is a low-complexity convolutional neural network with only 9,442 tunable parameters, that achieves a mapping of $\mathcal{F}_{\mathbf{P}}\left(\mathbf{H^{LS}}\right)$. A primary predictor, denoted as \( \mathcal{F}_{\mathbf{P}_{\mathrm{HDS}}}(\cdot) \) with a certain set parameters $P_{\mathrm{HDS}}$, is implemented using an InterpolateNet trained under high delay spread (HDS) conditions. It produces an enhanced estimate as: 
\begin{equation}
\mathbf{H}^{\text{DNN}} = \mathcal{F}_{\mathbf{P}_{\mathrm{HDS}}} \left( \mathbf{H}^{\text{LS}} \right),
\end{equation}
where \( \mathbf{H}^{\text{DNN}} \in \mathbb{C}^{N_f}\) is the InterpolateNet channel estimate. To assess the channel's temporal dispersion, we compute the IFFT of \( \mathbf{H}^{\text{DNN}} \), yielding a time-domain estimate to estimate the power delay profile (PDP). A classification function \( \Psi(\cdot) \) is then applied to compare \( \mathbf{h}^{\text{DNN}} \) with pre-defined PDP templates \( \mathbf{h}_{\mathrm{LDS}} \) and \( \mathbf{h}_{\mathrm{HDS}} \), resulting in the selection:
\begin{equation}
\ell^\star = \Psi\left( \mathbf{h}^{\text{DNN}}; \mathbf{h}_{\mathrm{LDS}}, \mathbf{h}_{\mathrm{MDS}} \right) \in \{ \text{LDS}, \text{MDS}, \text{HDS} \}.
\end{equation}

This index \( \ell^\star \) determines the most suitable neural network for final estimation. The deployed classification in this paper is a hard-decision process that directly compare the $\mathbf{H}^{\text{DNN}} $ with $\mathbf{h}_{\mathrm{LDS}}, \mathbf{h}_{\mathrm{MDS}}$. For each scenario, a dedicated InterpolateNet is trained using channel samples with corresponding power delay profile (PDP) characteristics. Specifically: 
\begin{itemize}
    \item \( \mathcal{F}_{\mathbf{P}_{\mathrm{LDS}}} \): optical channels with a sampled power delay profile (PDP) below $\mathbf{h_{\mathrm{LDS}}}$, 
    \item \( \mathcal{F}_{\mathbf{P}_{\mathrm{MDS}}} \): trained on optical channels with a sampled power delay profile (PDP) between LDS and HDS scenarios, 
    \item \( \mathcal{F}_{\mathbf{P}_{\mathrm{HDS}}} \): optical channels with a sampled power delay profile (PDP) of $\mathbf{h_{\mathrm{HDS}}}$.
\end{itemize}

The choice of the values of $\mathbf{h}_{\mathrm{LDS}}$ and $\mathbf{h}_{\mathrm{HDS}}$ is based on clustering environment measurement. As shown in \cite{luan2023achieving}, the HDS-trained InterpolateNet (also called HDS-Net) exhibits strong generalization across diverse scenarios and provide a good estimation for them. Therefore, as described in Algorithm.~\ref{Proposed method}, our framework initially invokes HDS InterpolateNet and use the result from this neural network to estimate the channel impulse response. This paper predicts the channel impulse response by using the IFFT operation and $\mathbf{H^{DNN}}$ from the first step. The proposed algorithm then compares $\mathbf{h_{\mathrm{HDS}}}$ and $\mathbf{h_{\mathrm{LDS}}}$ and only switches to LDS or MDS networks when the estimated delay profile significantly deviates from the HDS reference. This design ensures high estimation accuracy while maintaining computational efficiency, which has a maximum time complexity of $\mathcal{O}(N_f N_s) + \mathcal{O}(N_f \log N_f)$. 
\begin{algorithm}[H]
\caption{InterpolateNets Selection\&Prediction}
\label{Proposed method}
\begin{algorithmic}[1]
\Require \( \mathbf{h_{\mathrm{HDS}}}, \mathbf{h_{\mathrm{LDS}}}, \mathbf{P_\text{HDS}}, \mathbf{P_\text{MDS}}, \mathbf{P_\text{LDS}}, \mathcal{F} \)
\Ensure \(\forall i \in [1:L_{\mathrm{CP}}],\ |\mathbf{h_{\mathrm{HDS}}}(i)| > |\mathbf{h_{\mathrm{LDS}}}(i)| \)

\State \( \mathbf{H^{\mathrm{DNN}}} \gets \mathcal{F}_{\mathbf{P_\text{HDS}}}\left(\mathbf{H^{\mathrm{LS}}}\right) \)
\State \( \mathbf{h_{\mathrm{est}}} \gets \mathrm{Estimate \ time \ impulse \ response \ by \ } \mathbf{H_{\mathrm{DNN}}} \)
\State \( \mathbf{h_{\mathrm{abs}}} \gets \mathrm{abs}\left(\mathbf{h_{\mathrm{est}}}\right) \)

\If{\( \forall i \in [1:L_{\mathrm{CP}}], \mathbf{h_{\mathrm{abs}}}\left(i\right) < |\mathbf{h_{\mathrm{LDS}}}\left(i\right)| \)}
  \State \( \mathbf{H^{\mathrm{DNN}}} \gets \mathcal{F}_{\mathbf{P_\text{LDS}}}\left(\mathbf{H^{\mathrm{LS}}}\right) \)
\ElsIf{\( \forall i \in [1:L_{\mathrm{CP}}], \mathbf{h_{\mathrm{abs}}}\left(i\right) < |\mathbf{h_{\mathrm{HDS}}}\left(i\right)| \)} 
  \State \( \mathbf{H^{\mathrm{DNN}}} \gets \mathcal{F}_{\mathbf{P_\text{MDS}}}\left(\mathbf{H^{\mathrm{LS}}}\right) \) 
\Else
  \State \textbf{Do Nothing} 
\EndIf
\end{algorithmic}
\end{algorithm}
\section{Simulation Results}
\label{Simulation Results}
For the rest of this paper, $M$ is assumed to be 2 which represents a two-path channel for common indoor scenarios. Normalized mean square error (NMSE) is a key performance metric that evaluates the distance between the actual channel and the estimate of channel in the slot, which is defined as 
\begin{equation}
    \mathrm{NMSE}(\mathbf{\hat{H}_{Slot}}, \mathbf{H}) = \frac{\mathbb{E}\left\{\left\Arrowvert\mathbf{\hat{H}_{\mathrm{Slot}}} - \mathbf{H}\right\Arrowvert_{F}^{2}\right\}}{{\mathbb{E}\{\left\Arrowvert \mathbf{H}\right\Arrowvert_{F}^{2}\}}}, 
\end{equation}
where $\left\Arrowvert \cdot \right\Arrowvert_{F}^{2}$ is the Frobenius norm. The bit error ratio (BER) is another performance metric. 
\begin{table}[h]
    \renewcommand{\arraystretch}{1.2}
    \centering
    \caption{System settings for optical wireless channel}
    \label{simulation optical channel}
    \resizebox{\columnwidth}{!}{
    \begin{tabular}{p{0.25\textwidth} p{0.25\textwidth}} 
        \toprule
        \textbf{Parameters} & \textbf{Values} \\
        \midrule
        Room size & 5 m $\times$ 5 m $\times$ 5 m \\
        Number of Transmitter/Receiver & 1 $\times$ 1 \\
        Reflection coefficients ($\alpha$) & 0.7\\ 
        Transmitter FOV ($\Phi_{1/2}$) & 45$^\circ$ \\
        Receiver random rotation angles & 0$^\circ$ to 360$^\circ$ \\
        Receiver elevation angle & 0$^\circ$ to 30$^\circ$ \\ 
        Receiver FOV ($\phi_{1/2}$)& 45$^\circ$ \\
        Environment SNR & from 15dB to 30dB \\
        Collection area ($A$) & 1 cm$^2$ \\ 
        \bottomrule
    \end{tabular}}
\end{table}
The offline training dataset for the three InterpolateNet networks (LDS, MDS or HDS channels) consists of 100,000 channel samples, 95\% for training and 5\% for validation. Referring to the environment parameters given in Table.~\ref{simulation optical channel}, we set $\mathbf{h_{\mathrm{LDS}}}$ = 1e-4 $\times$ [6.4, 0.21930, 0.09676, 0.06175, 0.04517] and $\mathbf{h_{\mathrm{HDS}}}$ = 1e-4 $\times$ [5.5, 0.30126, 0.13441, 0.08609, 0.06310] respectively for this indoor scenario. We use MSE loss for training InterpolateNets with Adam optimizer. The initial learning rate is 0.0002 and is reduced by 0.3 for every 10 epochs, for up to 100 epochs. The mini-batch size is 64 and L2 regularization item is $10^{-9}$. 
\subsection{NMSE performance on randomly generated channels}
We first evaluate the NMSE performance on randomly generated channels over both signal-to-noise ratio (SNR) and time, to investigate the robustness to allowed SNRs. The test channels are generated randomly according to Table.~\ref{simulation optical channel}. 
\begin{figure*}[htbp]
\centering
\subfloat[\textnormal{NMSE performance over SNR from 15dB to 30dB.} \label{MSE_over_SNR}]{%
       \includegraphics[width=0.4\linewidth]{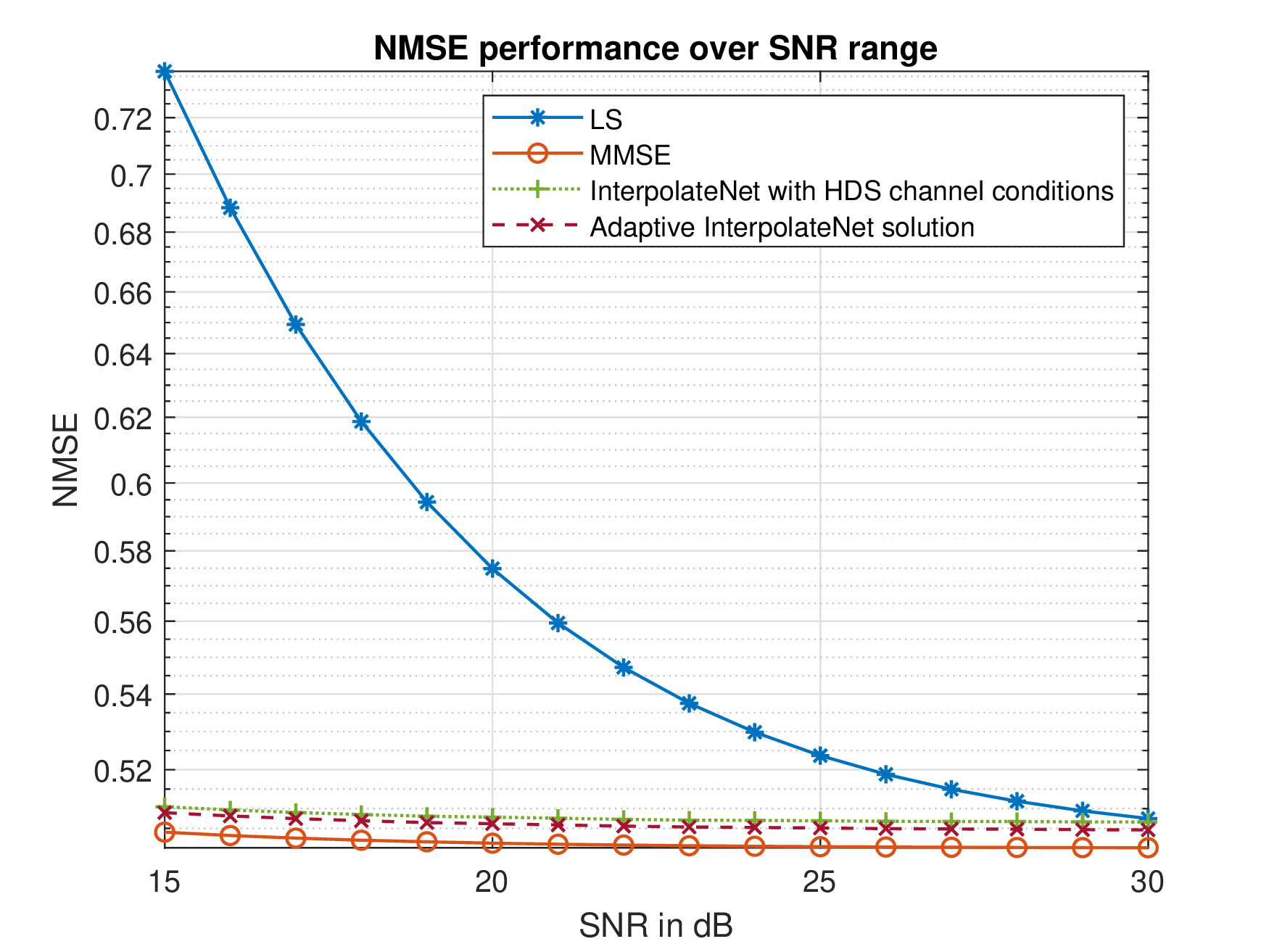}}
\hfill
\subfloat[\textnormal{NMSE performance over 90 seconds time (20dB SNR).} \label{MSE_over_time}]{%
        \includegraphics[width=0.4\linewidth]{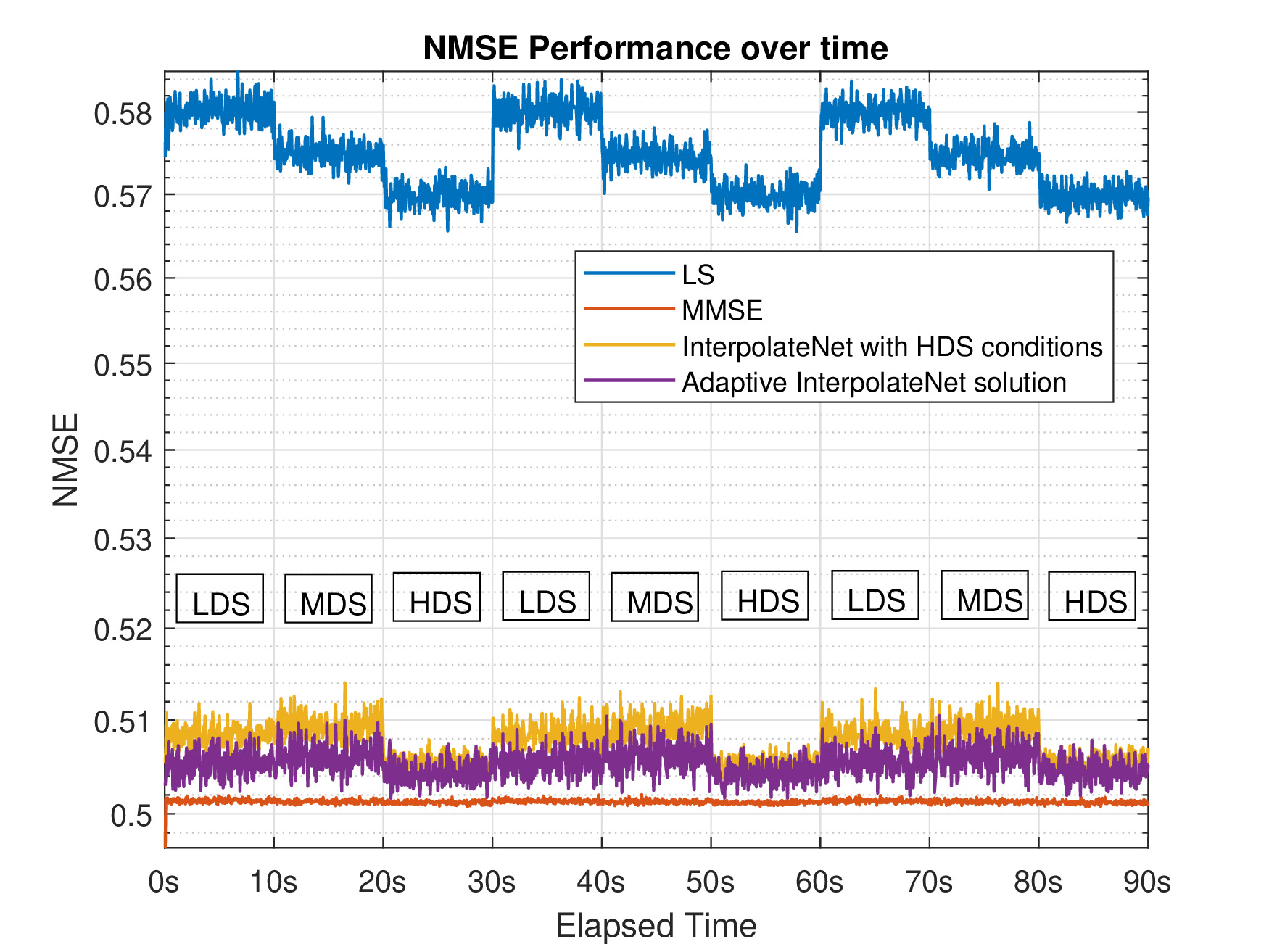}}
\caption{NMSE performance of LS, MMSE, InterpolateNet trained with HDS conditions and the proposed method. }
\end{figure*}

Figure.~\ref{MSE_over_SNR} shows the NMSE performance over the SNR from 15dB to 30dB and each sample is averaged by 100,000 independent channel realizations. The proposed method is shown to have superior performance than other methods except for MMSE method. At 30dB SNR, the adaptive InterpolateNet achieves a NMSE of 0.5046 which is 0.0020 lower than that of HDS InterpolateNet. The MMSE method has a minimum NMSE value of 0.5001. For figure.~\ref{MSE_over_time}, the channel will alter among LDS, MDS and HDS for each 10 seconds and each sample is averaged over 100 independent channel realizations. Both the proposed method and the InterpolateNet trained with HDS condition outperform the LS estimate but are worse than MMSE method. The proposed method provides a better NMSE performance with a mean value of 0.5034, which is lower than 0.5052 for the InterpolateNet trained with HDS condition. Compared to the MMSE estimate with NMSE of 0.5012, the proposed solution does not require any actual channel information for online implementation resulting in poorer estimate performance and the NMSE variance of the proposed method is reasonably small over time. 
\subsection{BER performance on randomly generated channels}
We also evaluate the BER performance of each method on randomly generated channels over SNR. 
\begin{figure}[htbp]
\centerline{\includegraphics[width=0.4\textwidth]{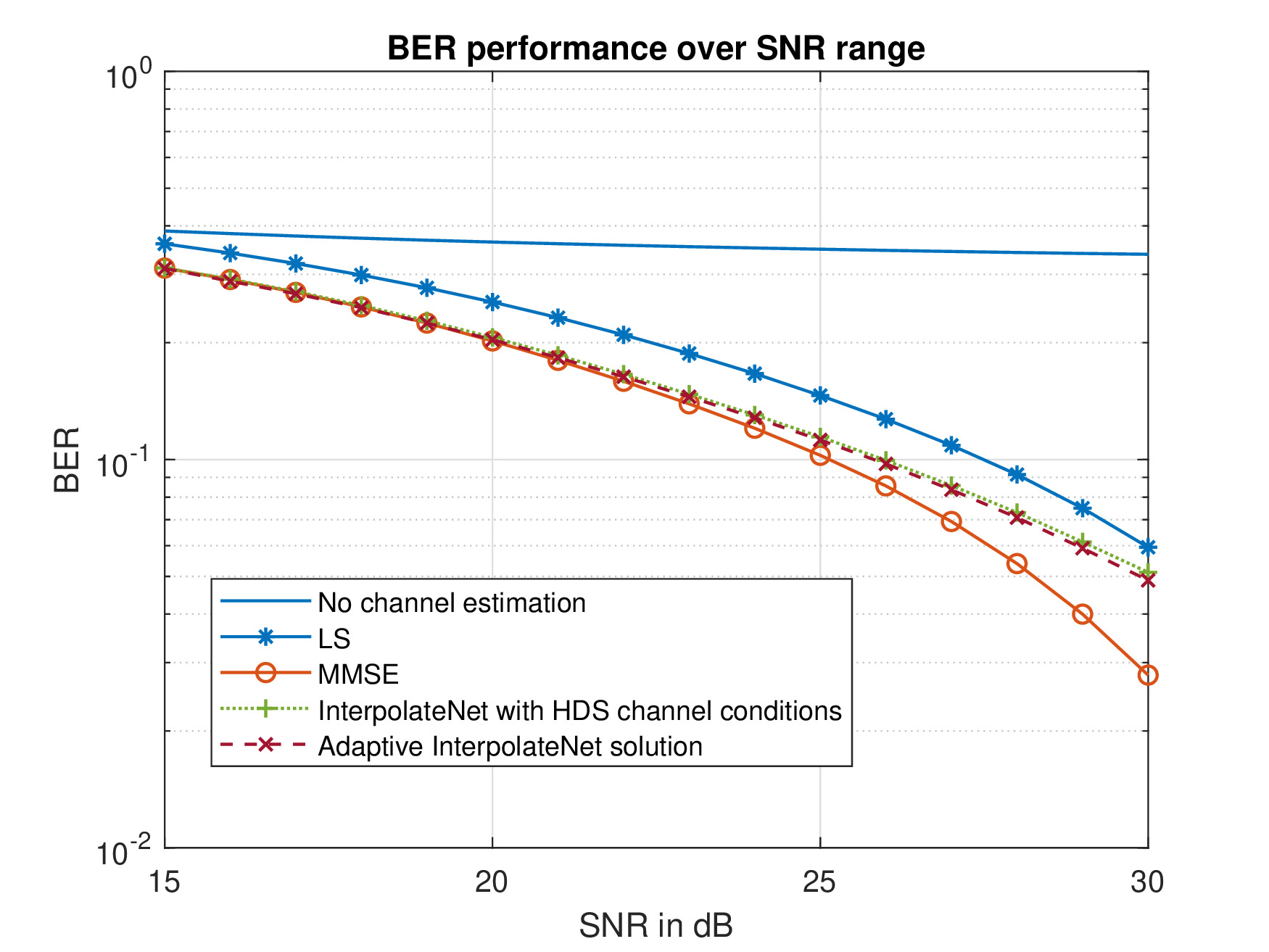}}
\caption{BER results of LS, MMSE, direct signal detection, HDS-Net and the proposed method over SNR from 15dB to 30dB.}
\label{BER_over_SNR}
\end{figure}

Figure.~\ref{BER_over_SNR} shows the BER performance over the SNR from 15dB to 30dB and each sample is averaged by 100,000 independent channel realizations. The proposed method also outperforms other methods except for MMSE method. At 30dB SNR, the adaptive InterpolateNet achieves a BER of 4.8\% while HDS InterpolateNet has a BER of 5.12\%. The MMSE method has a minimum NMSE value of 2.7\%. It should be noted that BER raises to 33.79\% for optical receivers that directly demodulate the received signals. Replacing InterpolateNet with high-complexity neural networks should further improve both the NMSE and BER performance. This is because the neural network capacity is small due to the low complexity property of InterpolateNet and the deployed optical wireless channels are weakly frequency-selective type. 
\section{Conclusion}
\label{Conclusion}
This paper proposes an adaptive low-complexity neural network solution to robustly estimate multi-path optical wireless channels. This proposed method selects one suitable candidate neural network based on the pre-estimate to provide an accurate prediction. From the simulation results, the proposed offline-trained solution performs robustly over randomly-generated optical channels and achieves better performance on both NMSE and BER than other methods which have no prior channel information known. Compared to the HDS-InterpolateNet and the LS estimate, the adaptive solution achieves reductions in NMSE and BER of 1.04\% and 1.5\% and 11.71\% and 20\% at 20 dB SNR respectively. 
\section*{Acknowledgments}
This research is supported by EPSRC projects EP/X04047X/1 and EP/Y037243/1. 
\ifCLASSOPTIONcaptionsoff
  \newpage
\fi

\bibliographystyle{IEEEtran}

\bibliography{Reference}

\end{document}